# A transformative superconducting magnet technology for fields well above 30 T using isotropic round wire multifilament $Bi_2Sr_2CaCu_2O_{8-x}$ conductor

D. C. Larbalestier[1], J. Jiang[1], U. P. Trociewitz[1], F. Kametani[1], C. Scheuerlein[2], M. Dalban-Canassy[1], M. Matras[1], P. Chen[1], N. C. Craig[1], P. J. Lee[1] and E. E. Hellstrom[1]

[1]Applied Superconductivity Center, National High Magnetic Field Laboratory, Florida State University, 2031 East Paul Dirac Drive, Tallahassee, FL 32310 USA

[2] European Organization for Nuclear Research, CH-1211 Geneva, Switzerland

April 9, 2013

We report here that magnetic fields of almost 34 T, far above the upper 24 T limit of $Nb_3Sn$, can be generated using a multifilament round wire conductor made of the high temperature cuprate superconductor $Bi_2Sr_2CaCu_2O_{8-x}$ (Bi-2212). A remarkable attribute of this Bi-2212 conductor is that it does not exhibit macroscopic texture and contains many high angle grain boundaries but nevertheless attains very high superconducting critical current densities $J_c$ of 2500 A/mm$^2$ at 20 T and 4.2 K. This Bi-2212 conductor does not possess the extreme texture that high $J_c$ coated conductors of $REBa_2Cu_3O_{7-x}$ (REBCO) require, avoiding also its high aspect ratio, large superconducting anisotropy and the inherent sensitivity to defects of a single filament conductor. Bi-2212 wires can be wound or cabled into almost any type of superconducting magnet and will be especially valuable for very high field NMR magnets beyond the present 1 GHz proton resonance limit of $Nb_3Sn$ technology. This demonstration that grain boundary limits to high $J_c$ can be practically overcome suggests the huge value of a renewed focus on grain boundary properties in non-ideal geometries, especially with the goal of translating the lessons of this Bi-2212 conductor into fabrication of multifilament round wire REBCO or Fe-based superconductors.

Any conductor for superconducting applications must develop high critical current density, $J_c$, in long length form. The difficulty of achieving the huge application dreams of the early 1990s for the cuprate high temperature superconductors (HTS) has very largely revolved around the complexity of making high $J_c$ in polycrystalline conductor forms, because of the huge current blocking effects of randomly oriented grain boundaries[1,2]. The principal solution pursued up until now has been to try to evade grain boundaries. Early on it was seen that a passage through the melt phase, first of YBCO[3], later of Bi-2212[4] could produce a self-organized, local grain alignment with much higher $J_c$ than was possible in randomly oriented polycrystals. There was particular interest in Bi-2212 because it could be conveniently melted inside an Ag-alloy sheath and for a time it seemed that Bi-2212 might develop into a viable conductor technology[5]. But its critical temperature ($T_c$) of only 90-95 K and its large superconducting anisotropy restricted its $J_c$ at 77 K and it was soon superseded by its sibling $(Bi,Pb)_2Sr_2Ca_2Cu_3O_{10-x}$ (Bi-2223), which, with a $T_c$ of ~110 K, did allow operation at 77 K[6]. High $J_c$ in Bi-2223 requires a complex reaction and deformation-induced texture process, many details of which still remain proprietary, whose goal is to minimize the density of high angle grain boundaries (HAGBs)[7]. After 10-15 years of development and the addition of a hydrostatic pressure step to close voids left behind during the Bi-2223 formation reaction (Bi-2212 + mixed oxides = Bi-2223), a uniaxial texture of ~15° FWHM has been achieved which allows a $J_c$ of order 500 A/mm² at 77K, self-field[8,9]. Although many aspects of a superconducting electrotechnology have been demonstrated successfully with Bi-2223, the real threshold for compelling, cost-effective applications typically occurs when $J_c$ exceeds 1000 A/mm² in strong fields, which explains why attention has increasingly turned to so called coated conductors (CC) based on $YBa_2Cu_3O_{7-x}$ (YBCO or now more generally REBCO RE = rare earth) in which a much better biaxial texture of 2-5° FWHM and much higher $J_c(77 K , sf)$ ~3 x10⁴ A/mm² can be obtained[10]. The great process complexity of the CC fabrication processes, however, is expensive and Bi-2223 still finds a market. A complication for both conductors is that their high costs, restricted lengths and large superconducting and shape anisotropies are all less than ideal for a versatile conductor, which should more ideally be round and multifilament, as is the case for the low temperature superconducting (LTS) wires, Nb-Ti and Nb₃Sn. In this respect therefore it is not surprising that HTS conductors have had great difficulty in displacing LTS conductors from applications such as MRI and NMR magnets, particle accelerators like the Tevatron and the Large Hadron Collider (LHC), or fusion devices like ITER. This may appear surprising since these applications all occur at liquid helium temperatures where the 5 or even 10 times higher $T_c$ values of HTS materials should allow their 4.2 K $J_c$ values to greatly exceed those of LTS conductors. Figure 1 compares the superconducting layer critical current densities $J_c$ of HTS and LTS conductors at standard operating temperatures of 4.2 or 1.9 K. Indeed REBCO has the highest $J_c$ by over an order of magnitude and an apparent advantage over both Bi-2212 and Bi-2223. The relatively poor $J_c(H)$ characteristics of Nb-Ti and Nb₃Sn are clear: although their $J_c$ values start off high, they fall below 1000 A/mm² at 11 T (at 1.9 K) for Nb-Ti, while Nb₃Sn crosses this threshold at about 17-18 T at 4.2 K or ~20 T at 1.9 K. This rapid $J_c$ drop occurs because of their low upper critical fields $H_{c2}$ of 15 and 30 T, respectively, while the flat $J_c(H)$ characteristics of the HTS conductors are due to their $H_{c2}$ exceeding 100 T[11]. Clearly high values of $J_c(H)$ do not by themselves make a competitive conductor technology.

Figure 1 shows thumbnail cross-sections of these state-of-the-art conductors. It is very noticeable that two of the HTS conductors, REBCO and Bi-2223 have highly aspected rectangular tape geometry, while all others are round wires. Figure 2 contrasts the strikingly different architectures of Nb-Ti and REBCO coated conductors (CC). The Nb-Ti Large Hadron Collider conductor contains more than 6000 filaments, each 6 μm in diameter and in intimate contact with high purity Cu that provides both electromagnetic stabilization against flux jumps and magnet quench protection[12]. The internal structure of the REBCO CC cannot be clearly seen at this scale: the support is a high strength, highly resistive Hastelloy substrate on which an oxide template and REBCO is deposited, the whole being surrounded by a 2 μm thick protective Ag layer on which a 20-50 μm thick high purity Cu layer is electroplated[10]. The advantages of this architecture for solenoid winding are considerable[13,14] because the dominant tensile hoop stresses are well supported by the very strong Hastelloy. But the disadvantages of the geometry are considerable too. Large magnets typically demand operation at many kA that require cables assembled from multiple strands. Due to their geometrical anisotropy, rectangular conductors are hard to cable except in special configurations[15,16,17]. The large superconducting anisotropy of strongly textured tape conductors (the *c*-axis is oriented perpendicular to the tape plane and the *a* and *b*-axes lie in the plane of the tape) is also a significant complication, as is the fact that the superconducting layer of REBCO coated conductor acts as one filament which makes it vulnerable to any defect that locally interrupts current flow, especially in a magnet, where its large stored energy can dissipate locally at any defect during quench. For all of these reasons a round wire, multifilament geometry with multiple independent current paths in which each superconducting filament is directly bonded to high conductivity normal metal is greatly preferred. Nb-Ti possesses all of these advantages, which is certainly an important reason why it is still by far the most widely produced superconductor, in spite of its $T_c$ being only 9 K and full use of its capabilities often, as in the LHC, requires operation at 1.9 K, not even 4.2 K[18]. Thus some of the great challenges that any HTS conductor should address are the need to be round, to be multifilamentary, to possess high $J_c$ and have high conductivity normal metal intimately bonded to the superconducting filaments. It is precisely the achievement of all these goals in Bi-2212 that we describe in this letter.

Figure 3a shows a cross-section of Bi-2212 round wire conductor in its as-drawn state, here embodied as a 0.8 mm diameter, with 666 ~15 μm diameter filaments embedded in a high purity Ag matrix with a strengthened Ag0.2%Mg outer sheath. The filaments are composed of Bi-2212 powder, which must be melted to establish continuous superconducting paths along each filament. During this melt and Bi-2212 regrowth Heat Treatment (HT), large grains of Bi-2212 form on cooling, making the plate-like filament structure shown in Figure 3b. A key characteristic of any powder-in-tube (PIT) process is that the powder cannot be 100% dense, because deformation of the conductor by wire drawing requires particle sliding inside the metal tubes. The key to our breakthrough to high $J_c$ was to understand that the crucial current limiting mechanism is not due to supercurrent blockage at high angle grain boundaries (HAGBs), as seemed entirely plausible given the need for high texture in Bi-2223 and REBCO1,2 but blockage by the agglomeration of the residual 30-40% uniformly distributed void or gas space into filament-diameter bubbles on melting the Bi-2212 powder[19]. These bubbles are seen very clearly in Figure 3c in samples quenched from the melt step of the HT. After full HT these bubbles are not generally visible, because 10-40 μm long Bi-2212 grains grow across the bubbles, providing a long-range but strongly compromised connectivity that greatly reduces the long range $J_c$. Figures 3d and 3e show two 45° rotated topographic views that clearly expose this highly aspected Bi-2212 growth. Figure 3f provides crystallographic detail of a reacted filament in which some local texture is evident but where the presence of [001], [100], and [110] oriented grains and many HAGBs is clear. The key point of this paper is that such HAGBs do not prevent high $J_c$.

The effective solution to the problem of bubbles is to apply sufficient overpressure (OP) during HT to prevent their formation and the deleterious creep dedensification[20,21] that the internal gas pressure drives. Figure 4 shows that reaction under pressures up to 100 bar can raise the whole-conductor (or engineering) current density $J_E$ by up to 8 times. Although these samples were all short (~8 cm), they were reacted with closed ends so that gas generated internally was trapped in the wire as it would be in long lengths. Wire densities higher than 95% were obtained for >50 bar reactions. The effect of overpressure can be most effectively seen by x-ray tomography as shown in Figures 4b and c, where the extensive porosity and tendency for a wire processed at 1 bar to dedensify and leak contrasts (Fig. 4b) strongly with the rather uniform full density of a wire OP processed at 100 bar (Fig. 4c). Figure 4a shows that although the Bi-2212 occupies only 25% of the cross-section of the wire, $J_E$ reaches almost 1000 A/mm$^2$ at 4.2 K, 5 T ($J_c$ is ~ 4000 A/mm$^2$, higher even than the $J_c$ of optimized Nb-Ti at 4.2 K, 5 T).

Earlier experiments had showed us that high $J_c$ was possible in this system if we took very special care to allow gas to escape from short wire lengths[22], but the challenge of how to control the dedensification caused by internal gas pressure in long length wires remained. Developing a short length model of the long coil length wires was essential. Our supposition was that closing the wire ends would make the properties independent of length, but to prove that long samples performed as well as short samples, we made a test coil from 30 m of 1.4 mm dia. Bi-2212 wire, as shown in Figure 5. Here we were constrained to only 10 bar by our larger diameter OP furnace. Nevertheless the coil behaved excellently. In 31.2 T background field it generated 2.6 T, more than twice the field of an earlier similar sized coil reacted at 1 bar[13]. It was quenched multiple times without any damage and generated a total field of 33.8 T.

We also tested sections from the 8 cm long closed-end samples shown in Figure 4 at fields up to 31 T and multiple samples with properties corresponding to the data in Figure 1 and Figure 6 were obtained. What is now clear is that fully dense Bi-2212 wires for which dedensification has been prevented by overpressure processing have higher $J_c$ than the very best Bi-2223 tape samples[9] with considerable uniaxial texture, though $J_c$(Bi-2212) is still almost an order of magnitude lower than for REBCO tape with strong biaxial texture. However, this $J_c$ advantage of REBCO disappears when the whole-conductor current density $J_E$ is considered because the REBCO layer is only 1% of the whole conductor cross-section. Figure 6 shows that now 100 bar processed Bi-2212 (25% superconductor) has significantly higher $J_E$ than either Bi-2223 (40% superconductor) or REBCO coated conductor (1% superconductor). The $J_E$ values attained by Bi-2212 ($J_E$ 700, 630 and 500 A/mm$^2$ at 15T, 20T and 30T) after OP HT are all high enough for winding very high field solenoids for very high field NMR or for 20 T upgrades for the LHC[23] or other high field magnet uses.

There are multiple broad implications of this work. First is that overpressure processing for Bi-2212 generates a conductor with very broad applications for high field magnet use well beyond the capabilities of Nb$_3$Sn and Nb-Ti. The round wire, multifilament architecture of Bi-2212 is highly advantageous compared to the essentially single architecture, wide-tape forms in which Bi-2223 and REBCO CC are manufactured. Moreover, multiple filament architectures of Bi-2212 are easily possible because optimization of the superconducting properties occurs after working it to final size, not during manufacture as is the case for Bi-2223 and REBCO. The round wire architecture means that winding high homogeneity coils such as those needed for NMR is much easier and the magnetization induced in low field regions of the windings is much less troublesome than is the case for tape conductors[24,25]. Accelerator use is likely to be much more feasible than for tape conductors since the standard form of transposed flat cable, the Rutherford cable, can be made as easily from Bi-2212 round wire as from $Nb_3Sn$ wire [26,27]. In the short run it would appear that the present world record for magnetic field generated by a superconducting coil (35.4 T in 31.2 T background, made with CC[14]) should be exceeded using a Bi-2212 coil processed at 100 bar. However, the long term implications of Figure 3f are even more powerful. The grain structure visible in this dense filament clearly contains many high angle grain boundaries, as well as residual liquid that has transformed largely to Bi-2201. Some of these defects and grain boundaries must be blocking current, reducing the long-range $J_c$ well below that within grains of Bi-2212 and also reducing $J_c$ well below that of biaxially textured REBCO where almost no high angle grain boundaries are present (Figure 1). But as Figure 6 emphasizes, Bi-2212 filaments are well enough connected that their 25% cross-section makes their overall current density $J_E$ the best of any available conductor above about 17 T. The really striking opportunity offered by Figure 3f appears when the inset of Figure 6 is considered. At present all of superconducting magnet technology is confined to the small arcs at lower left corresponding to the $H_{c2}(T)$ phase space of isotropic Nb-Ti and $Nb_3Sn$. Because of the huge electronic anisotropy of both Bi compounds, their irreversibility field $H_{irr}(T)$ has a quite different shape (see the inset of Figure 6), making neither conductor particularly valuable for magnets of more than 10 T above 10 K (2212) or 20 K (Bi-2223). But the much lower anisotropy REBCO shows an irreversibility field of 30 T at ~55 K. As present helium shortages[28] and recent price rises show, there would be great interest in a helium-free, cryocooler-driven superconducting magnet technology capable of 3-15 T operating in the 20-60 K temperature range. To access this range, we need to understand how to engineer the key properties of the Bi-2212 grain boundaries of Figure 3f into REBCO. Recent studies of *ex situ* $YBa_2Cu_3O_{7-x}$ [29] and $(Ba_{0.6}K_{0.4})Fe_2As_2$ [30] show that curved, "real" grain boundaries can often show much better transparency than more idealized, planar grain boundaries[2] and it is precisely these that appear to enable high $J_c$ with much less texture and the desirable round wire multifilament architecture presented here. A REBCO equivalent of the multifilamentary Bi-2212 round wire described here would be truly transformational for superconducting magnet technology.

## Methods

Wires were fabricated by Oxford Superconducting Technology. Wire ends were sealed by dipping in molten silver followed by either a standard heat treatment (HT) at 1 bar or an OP HT [20], using pure $O_2$ for 1 bar HT and $O_2$-Ar mixtures for OP HT. Field emission scanning electron microscope (FESEM) imaging and Orientation Image Analysis were performed in a Carl Zeiss 1540 XBCrossbeam® instrument using a high speed Hikari camera. X-ray tomograms were acquired using a monochromatic 70.0 keV x-ray beam with a bandwidth of 0.7 keV at the High Energy Scattering Beamline ID15A at the European Synchrotron Radiation Facility (ESRF), Grenoble using a high resolution imaging detector with a 15 µm-thick

LuAG:Ce$^{2+}$ scintillator that converts the x-ray absorption signal into visible light, which is then magnified and recorded by a high speed CCD camera with an image pixel area of 1.194 ×1.194 µm$^2$.

Transport critical currents ($I_c$) were measured on 45 mm long Bi-2212 wires at 4.2 K and fields up to 31 T applied perpendicular to current flow. $I_c$ was determined at 1µV/cm. The coil of Figure 5 was wound using 30 m of 1.4 mm diameter Bi-2212 wire with a ~13 µm thick TiO$_2$ (nGimat LLC) insulation layer, which is 10 times thinner than earlier ceramic braid insulations [15]. The coil was heat-treated in an OP furnace with a slightly too small homogeneous hot zone. The body of the coil was heat treated in a ±0.5ºC temperature zone, but coil performance was slightly degraded because the terminals extended into a ±4.5ºC temperature zone. Reaction of subsequent coils in more homogeneous furnaces with full 100 bar OP HT capability should allow generation of a new record high field for a superconducting coil significantly above 35 T.

**Acknowledgements**

This work was carried out within the Very High Field Magnet Collaboration (VHFSMC) which was supported by an ARRA grant of the US Department of Energy, Office of High Energy Physics, amplified by the National High Magnetic Field Laboratory, which is supported by the National Science Foundation under NSF/DMR-1157490 and by the State of Florida. We are grateful to many discussions from partners within the VHFSMC collaboration, especially Arup Ghosh (BNL), Arno Godeke (LBNL), Andrea Malagoli (Now CNR-SPIN, Genoa Italy), and Tengming Shen (Fermilab). We acknowledge the ESRF for beam time at ID15A.

**Author contributions**

JJ and MM reacted the samples, NCC, MD-C and UPT performed the transport critical current measurements, FK and JJ performed the metallography and EBSD, CS performed the x-ray tomography, MD-C, PC and UPT constructed and tested the coil and DCL, EEH and PJL led the work and took the lead in preparing the paper.

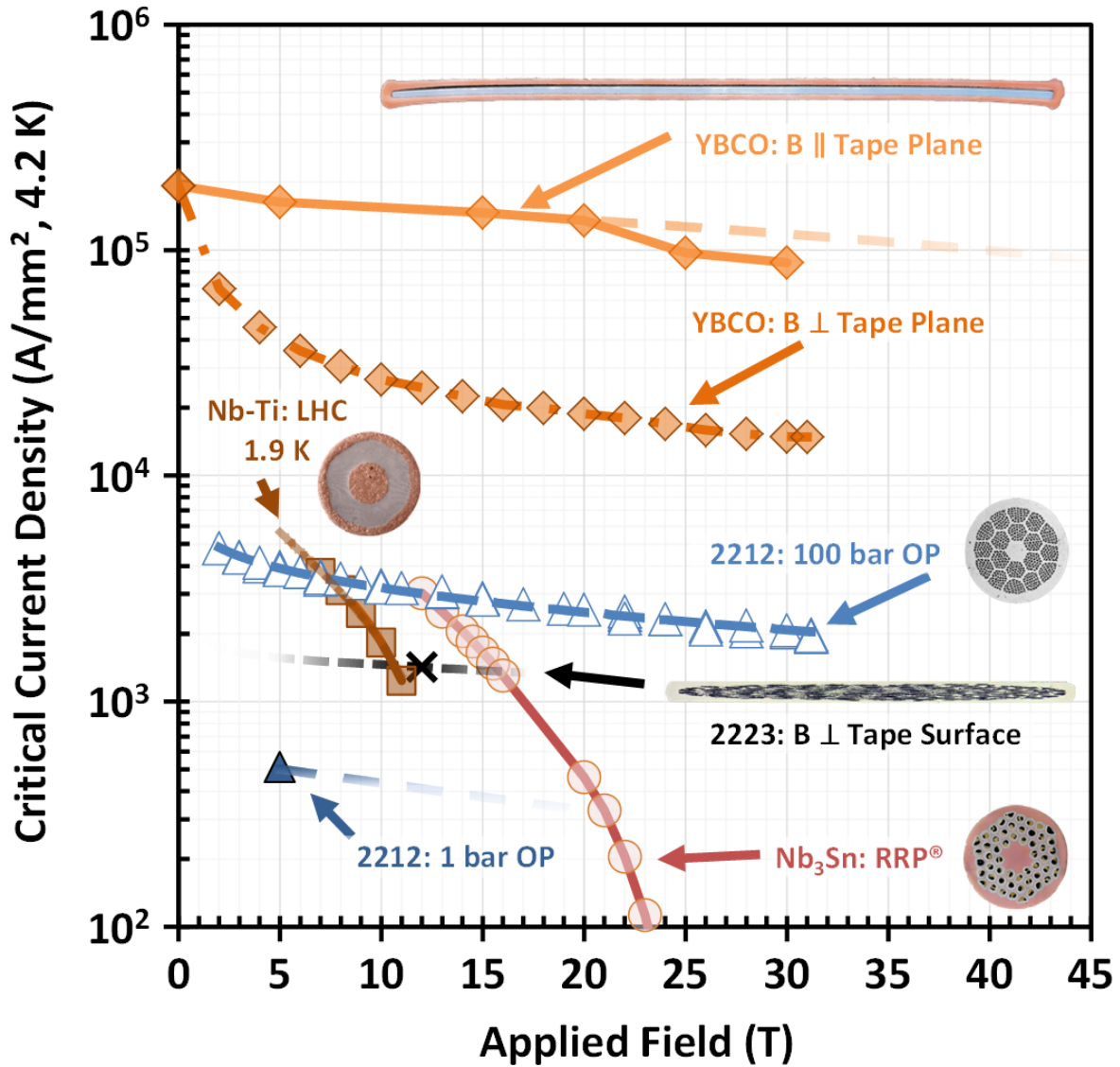

Figure 1. Critical current densities for representative state-of-the-art conductors used for superconducting magnet construction: $Nb_3Sn$ (Rod Restack Process RRP made by Oxford Superconducting Technology (OST)), round wire Bi-2212 (OST) reacted under total pressure of 1 and 100 bar at a fixed $O_2$ partial pressure of 1 bar, Bi-2223 (Sumitomo DI-BSCCO) and REBCO (SuperPower) coated conductor. All evaluations are at 4.2 K except for the LHC Nb-Ti strand which is used at 1.9 K and all were measured at the NHMFL except for the LHC Nb-Ti (courtesy T. Boutboul, CERN) and the Bi-2223 [courtesy K. Hayashi, Sumitomo]. In virtually all HTS magnet applications the coil performance is limited by the lower $J_c$ values of the anisotropic conductor, thus making the relevant $J_c$ curve that for H normal to the tape plane. $Nb_3Sn$ $J_c$ falls off rapidly above 20 T because its zero temperature upper critical field $H_{c2}(0)$ is only 30 T, compared to over 100 T for Bi-2212, Bi-2223 and YBCO. The relevant metric for conductor use must also consider the amount of superconductor that can be inserted into the conductor: this is at present only 1-2% for YBCO coated conductors, while it is 25-40% for Bi-2212 and Bi-2223 and up to 50% for Nb-Ti and $Nb_3Sn$. Figure 6 reflects these different superconductor fill factors in presenting the whole conductor, engineering current densities $J_E$.

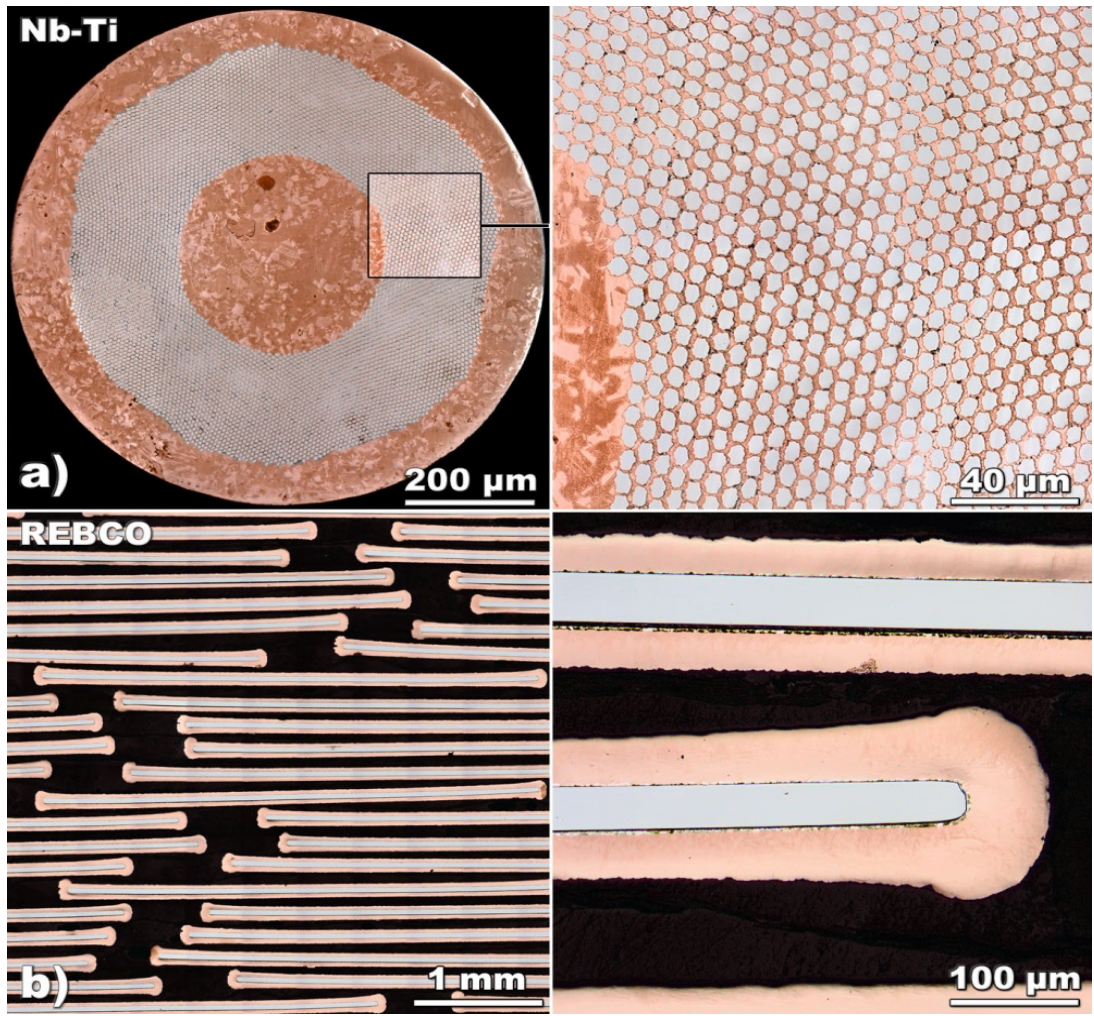

Figure 2: The extremes of present conductor technology: a.) CERN/LHC 0.825 mm diameter superconducting strand containing 6426 Nb-Ti filaments each 6 µm in diameter and each surrounded by pure Cu needed to confer electromagnetic stability against flux jumps and magnet quench protection when superconductivity is lost. b.) state-of-the-art REBCO coated conductor grown on a 50 µm thick strong Hastelloy substrate on which a 3 layer oxide seed and buffer layer with a strong [001] oriented biaxial texture (FWHM 2-3°) is developed by ion beam assisted deposition (IBAD). The ~1 µm thick REBCO layer was in this case deposited by MOCVD and protected by a 2 µm Ag layer. Finally Cu is electroplated onto the whole stack in thicknesses of 20-50 µm. A typical conductor is 4 mm wide and 0.1 µm thick, thus having an aspect ratio of 40. The right hand side of each image shows a magnified view of the conductors shown in full section at left. In this case the REBCO coated conductor has 50 µm thick Cu to provide full magnet quench protection rather than the standard 20 µm thickness that is used to calculate $J_E$ in Figure 6.

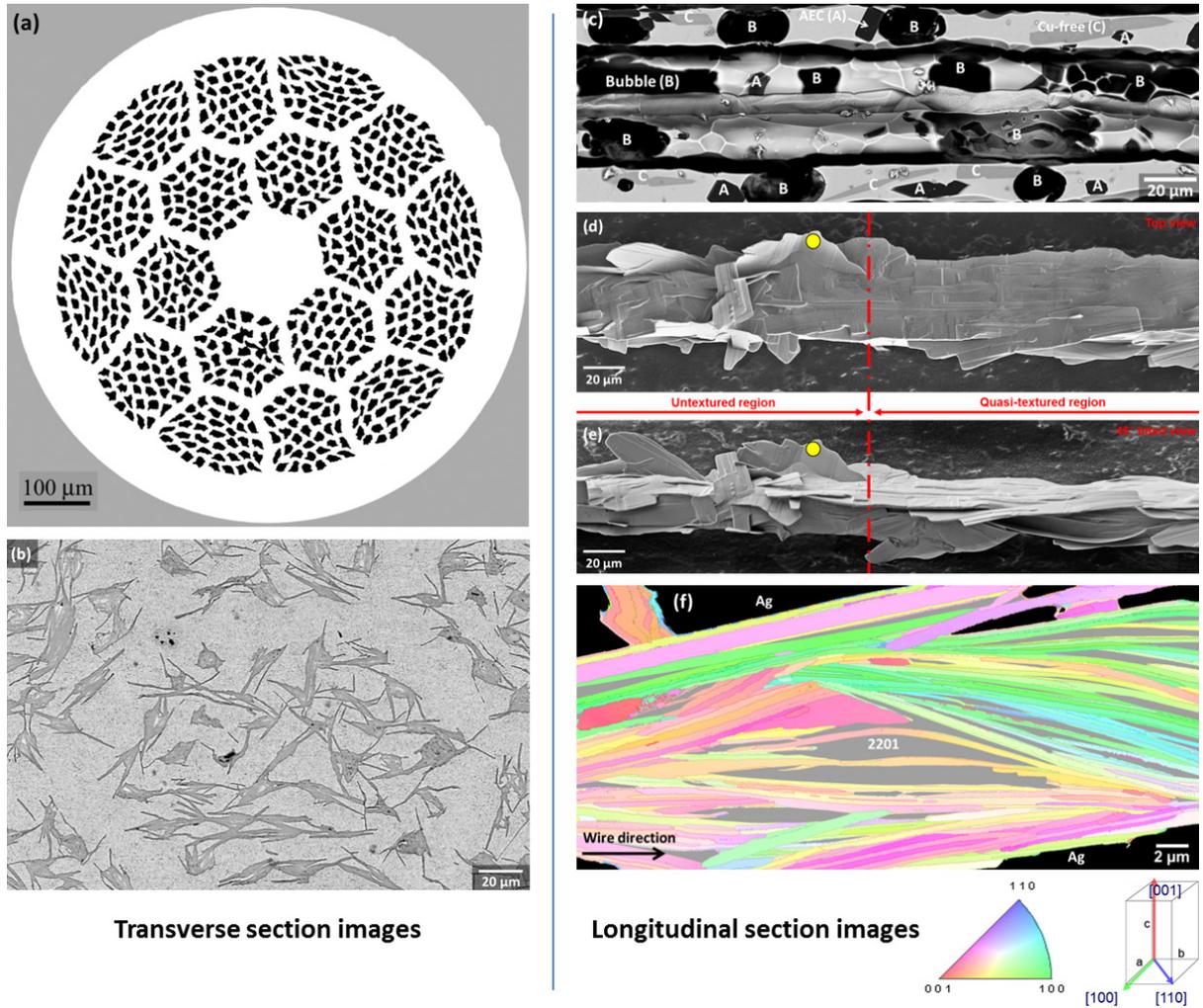

**Figure 3:** Macro- and microscopic images of the Bi-2212 conductor architecture: a.) Transverse section of an as-drawn 0.8 mm diameter wire composed of 18 x 37  15 μm diameter Bi-2212 powder filaments embedded in a Ag matrix before the melt process HT needed to establish long-range superconducting connectivity; b.) Transverse section centered on one 37 filament bundle showing the large aspected grains of 2212 that form after the HT; c.) Large agglomerated bubbles (B) are present in the filaments when the powder in a.) is melted; d.) and e.) Longitudinal views (rotated by 45°) of one fully-processed filament showing the highly-aspected 2212 grains formed during the HT. The yellow dot shows the same point on the filament in the two views; f.) Electron backscatter diffraction (EBSD) image of a polished dense filament section. The colors correspond to the orientation of the grains (shown below in f.). The EBSD image confirms both local grain-to-grain texture, and also shows multiple high angle GBs and residual liquid that has largely transformed to Bi-2201. These features do not however prevent development of high $J_c$. The filaments shown in c.) – e.) were exposed by etching away the surrounding Ag matrix.

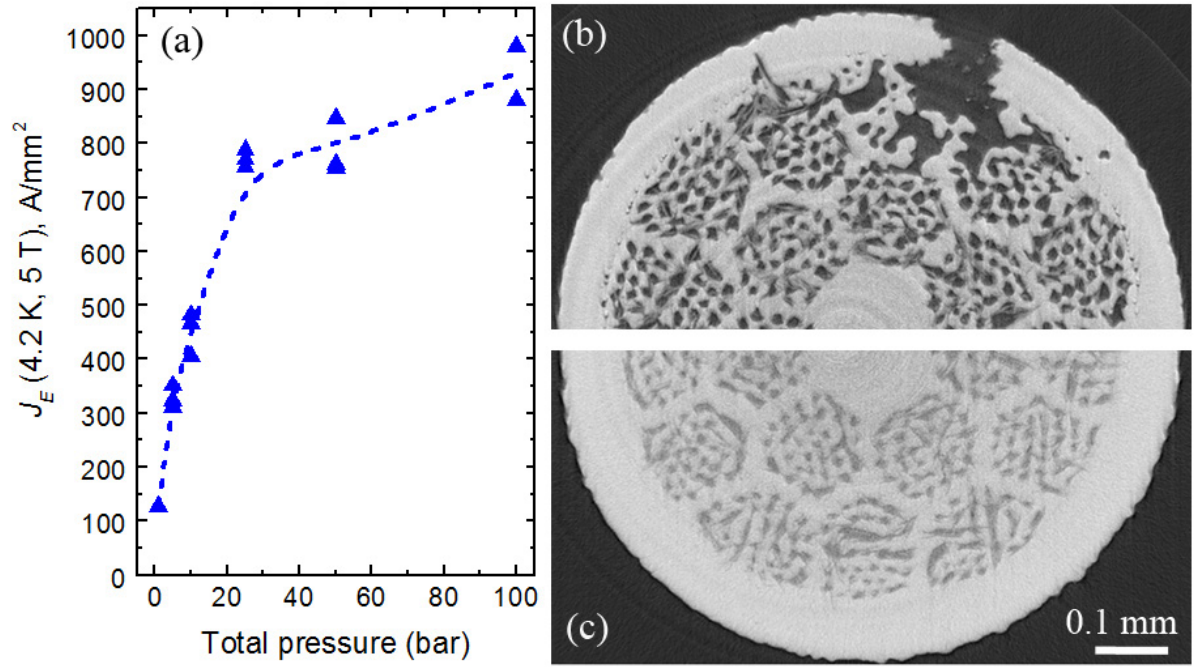

**Figure 4: a).** Overall conductor current density $J_E$(4.2K, 5T) as a function of the overpressure applied during HT to wires shown in Figure 3. The origin for the large increase in $J_E$ is evident by comparing Figures 4b) and 4c) where X-ray tomograms of 1 bar and 100 bar OP processed wires are shown. The extensive porosity of the final wire processed under 1 bar is clear as is the high density of the 100 bar processed wire. The 8 fold increase in $J_E$ is due to the elimination of leakage and suppression of the void porosity seen in Figure 4b by the 100 bar OP applied during HT. X-ray tomography was performed at the ID15A beamline in collaboration with M. Di Michiel, European Synchrotron Radiation Facility, Grenoble.

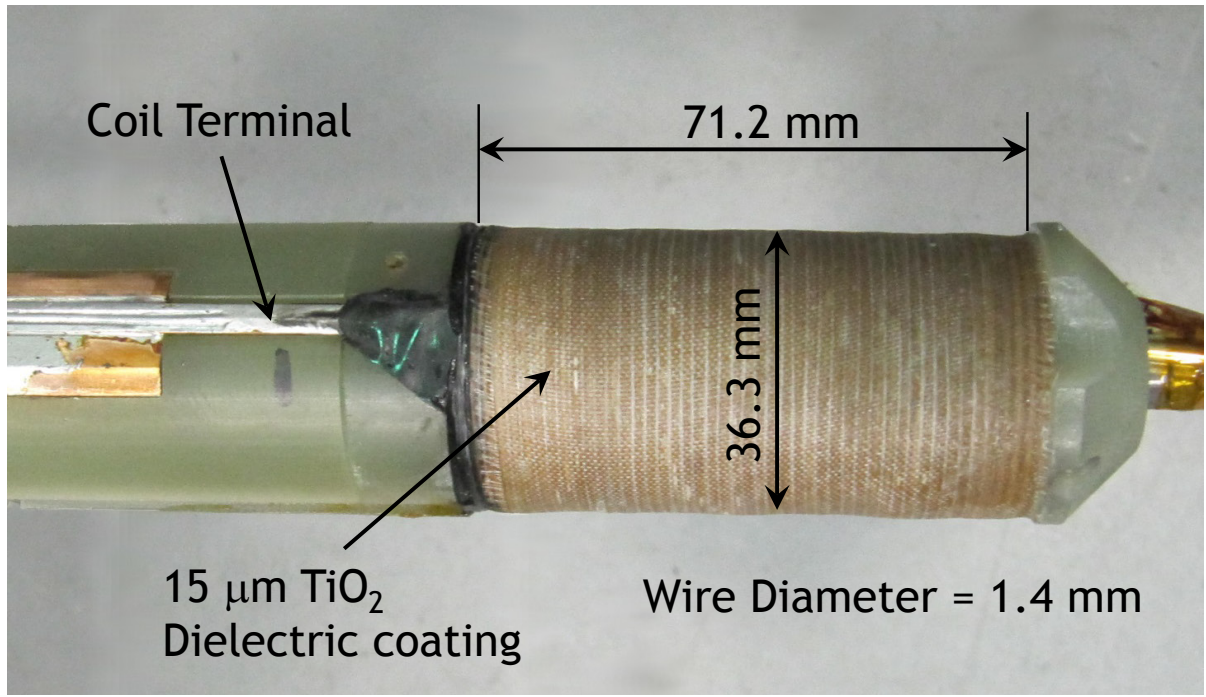

Figure 5: High field demonstration coil wound with 30 m of 1.4 mm diameter Bi-2212 wire. This coil was heat treated at 10 bar in an OP furnace. It operated in 31.2 T background field, generating a 2.6 T field increment for a total field of 33.8T at 6.7 mT/A. For comparison the previous coil heat treated in 1 bar only generated an additional field of 1.1 T[13]. The coil was quenched multiple times without damage. It achieved a maximum quench current of 388 A ($J_E$ = 252 A/mm$^2$ at 33.8 T).

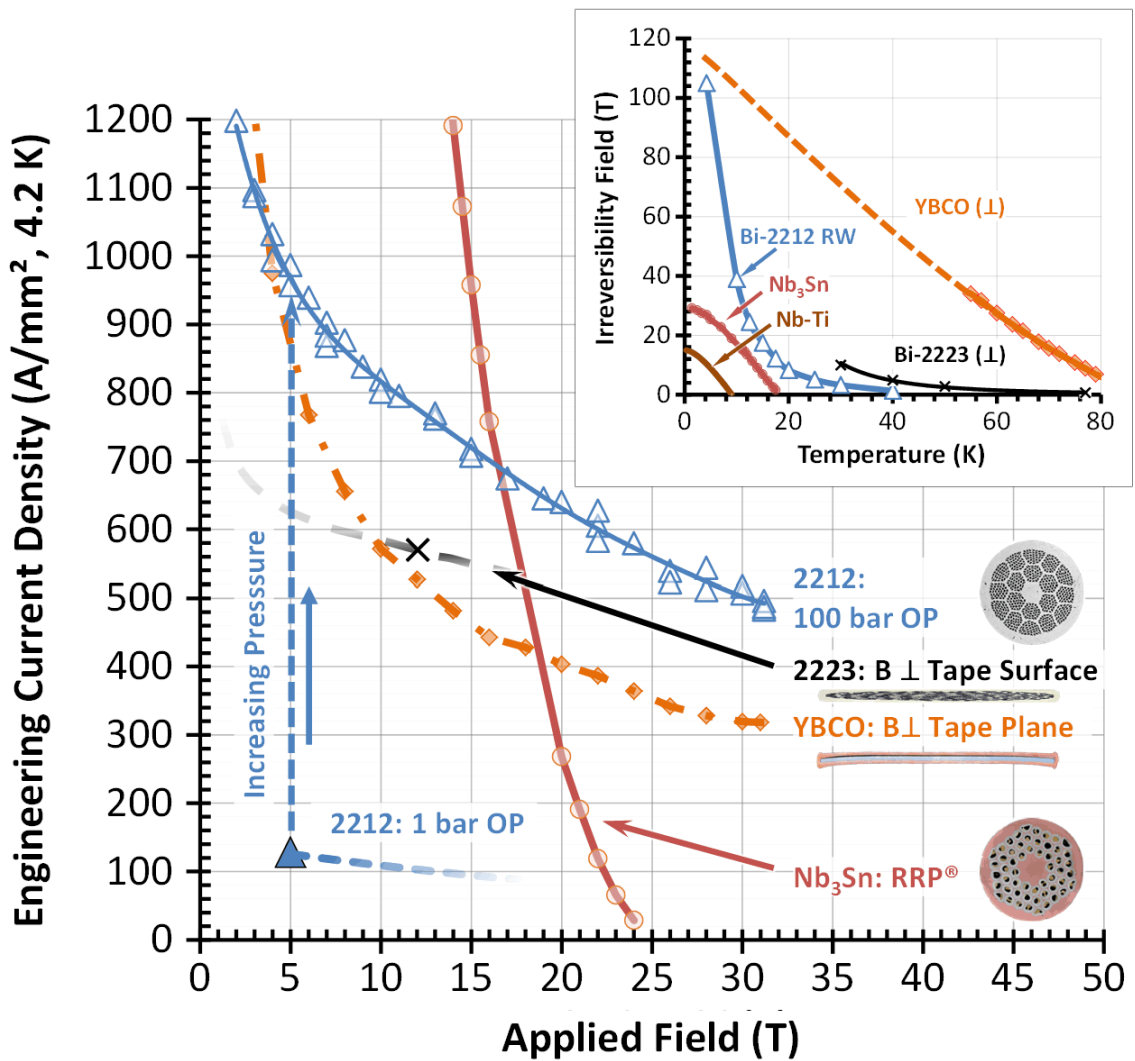

Figure 6: Overall conductor current densities $J_E$ possible for conductors made of REBCO, Bi-2223, Bi-2212 and Nb$_3$Sn used for constructing high field magnets generating up to 24 T (Nb$_3$Sn) and up to 35 T (REBCO). It is now seen that OP processed Bi-2212 has the highest conductor current density $J_E$ of any available conductor above 17 T and moreover that this is achieved in the most desired round-wire, multi-filament form. The inset shows upper critical field or irreversibility field data for the conductors. The great opportunities offered by REBCO at temperatures well above 4 K are clear.

**References**

[1] Larbalestier, D., Gurevich, A., Feldmann, D. M., Polyanskii, A. High-T-c superconducting materials for electric power applications. *Nature* **414**, 368-377 (2001).

[2] Hilgenkamp H., Mannhart J., Grain boundaries in high-Tc superconductors, *Rev. Mod. Phys.* **74**, 485-549 (2002)

[3] Jin, S., Tiefel, T. H., Sherwood, R. C., Kammlott, G. W., Zahurak, S. M. Fabrication of dense $Ba_2YCu_3O_{7-\delta}$ superconductor wire by molten oxide processing. *Appl. Phys. Lett.* **51**, 943-945 (1987).

[4] Heine, K., Tenbrink, J. and Thöner, M. High-field critical current densities in $Bi_2Sr_2Ca_1Cu_2O_{8+x}$/Ag wires. *Appl. Phys. Lett.* **55**, 2442-2443 (1989).

[5] Tenbrink, J., Wilhelm, M., Heine, K., Krauth, H. Development of high-Tc superconductor wires for magnet applications. *IEEE Transactions on Magnetics* **27**, 1239-1246 (1991).

[6] Sato, K. *et al.* High-Jc silver-sheathed Bi-based superconducting wires. *IEEE Transactions on Magnetics* **27**, 1231-1238 (1991).

[7] Rupich, M.W. & Hellstrom, E.E. (2012). Bi-Sr-Ca-Cu-O HTS Wire. In Rogalla, H. & Kes, P. P. H. (Eds.), *One Hundred Years of Superconductivity*. (pp. 671-689). CRC Press/Taylor & Francis Group.

[8] Yuan, Y. *et al.* Significantly enhanced critical current density in Ag-sheathed $(Bi,Pb)_2Sr_2Ca_2Cu_3O_x$ composite conductors prepared by overpressure processing in final heat treatment. *Appl. Phys. Lett.* **84**, 2127-2129 (2004).

[9] Kobayashi, S. *et al.* Controlled over pressure processing of Bi2223 long length wires., *IEEE Transactions on Applied Superconductivity* **15**, 2534-2537 (2005).

[10] Malozemoff, A.P. and Yamada, Y. (2012). Coated Conductor: Second Generation HTS Wire. In Rogalla, H. & Kes, P. P. H. (Eds.), *One Hundred Years of Superconductivity*. (pp. 689-702). CRC Press/Taylor & Francis Group.

[11] Sekitani T., Miura N., Ikeda S., Matsuda Y.H., Shiohara Y., Upper critical field for optimally-doped $YBa_2Cu_3O_{7-x}$, *Physica* B **346-347**, 319-324 (2004).

[12] M N Wilson, Superconducting Magnets. Clarendon Press, Oxford 1987.

[13] Weijers, H.W. *et al.* High Field Magnets with HTS Conductors. *IEEE Transactions on Applied Superconductivity*, **20**, 576-582 (2010).

[14] Trociewitz, U. P. *et al.* 35.4 T field generated using a layer-wound superconducting coil made of $(RE)Ba_2Cu_3O_{7-x}$ (RE = rare earth) coated conductor. *Appl. Phys. Lett.* **99**, 202506 (2011).

[15] Goldacker, W. *et al.*, Status of high transport current ROEBEL assembled coated conductor cables. *Supercond. Sci. Technol.* 22, 034003 (2009).

[16] Laan, D. C. van der, Lu, X. F. & Goodrich, L. F. Compact $GdBa_2Cu_3O_{7-\delta}$ coated conductor cables for electric power transmission and magnet applications. *Supercond. Sci. Technol.* **24,** 042001 (2011).

[17] Takayasu, M., Chiesa, L., Bromberg, L. & Minervini, J. V. HTS twisted stacked-tape cable conductor. *Supercond. Sci. Technol.* **25**, 014011 (2012).

[18] Rossi, L., Bottura, L., Superconducting Magnets for Particle Accelerators. *Reviews of Accelerator Science and Technology* **5**, 51-89 (2012)

[19] Kametani, F. *et al.*, Bubble formation within filaments of melt-processed Bi2212 wires and its strongly negative effect on the critical current density. *Supercond. Sci. Technol.* **24**, 075009 (2011).

[20] Malagoli A. *et al.,* Evidence for length-dependent wire expansion, filament dedensification and consequent degradation of critical current density in Ag-alloy sheathed Bi-2212 wires, to appear in *Supercond. Sci. Technol.* (2013).

[21] Shen T., Ghosh, A., Cooley, L., Jiang, J., Role of internal gases and creep of Ag in controlling the critical current density of Ag-sheathed $Bi_2Sr_2CaCu_2O_x$ wires, submitted to *J. of Applied Physics*.

[22] Jiang, J. *et al.* Doubled Critical Current Density in Bi-2212 Round Wires by Reduction of the Residual Bubble Density. *Supercond. Sci. Technol.* **24**, 082001 (2011).

[23] Bottura, L., De Rijk, G., Rossi, L. & Todesco, E. Advanced Accelerator Magnets for Upgrading the LHC. *IEEE Transactions on Applied Superconductivity* **22**, 4002008–4002008 (2012).

[24] Ahn, M. C. et al. Spatial and Temporal Variations of a Screening Current Induced Magnetic Field in a Double-Pancake HTS Insert of an LTS/HTS NMR Magnet. *IEEE Transactions on Applied Superconductivity* **19**, 2269 –2272 (2009).

[25] Yanagisawa, Y. et al. Magnitude of the Screening Field for YBCO Coils. *IEEE Transactions on Applied Superconductivity* **21**, 1640 –1643 (2011).

[26] Godeke, A. *et al.* Wind-and-react Bi-2212 coil development for accelerator magnets. *Supercond. Sci. Technol.* **23**, 034022 (2010)

[27] Barzi, E., Lombardo, V., Turrioni, D., Baca, F. J., Holesinger, T. G. BSCCO-2212 Wire and Cable Studies. *IEEE Transactions on Applied Superconductivity* **21**, 2335-2339 (2011).

[28] Cho A., NRC Urges U.S. to Rethink Sale of Helium Reserve. *Science* **327**, 511 (2010).

[29] Feldmann, D. M., Holesinger, T.G., Feenstra, R., Larbalestier, D. C. A review of the influence of grain boundary geometry on the electromagnetic properties of polycrystalline $YBa_2Cu_3O_{7-x}$ films. *Journal of the American Ceramic Society* **91**, 1869-1882 (2008).

[30] Weiss, J. D, *et al.* High intergrain critical current density in fine-grain $(Ba_{0.6}K_{0.4})Fe_2As_2$ wires and bulks. *Nature Materials* **11**, 682-685 (2012).